\documentclass{article}

\usepackage{microtype}
\usepackage{graphicx}
\usepackage{subfigure}
\usepackage{booktabs} 

\usepackage[utf8]{inputenc} 
\usepackage[T1]{fontenc}    
\usepackage{hyperref}       
\usepackage{url}            

\usepackage{booktabs}       
\usepackage{amsfonts}       
\usepackage{nicefrac}       
\usepackage{microtype}      
\usepackage{xcolor}
\usepackage{amsmath, amssymb}
\usepackage{graphicx}
\usepackage{wrapfig}
\usepackage{array}
\usepackage{float}
\usepackage{wrapfig}
\usepackage{comment}
\usepackage{caption}
\usepackage{epstopdf}
\usepackage{dblfloatfix} 
\usepackage{afterpage}

\usepackage{hyperref}

\usepackage[accepted]{icml2020}

\icmltitlerunning{Uncertainty-Aware Lookahead Factor Models}

\begin{document}

\twocolumn[
\icmltitle{Uncertainty-Aware Lookahead Factor Models 
for Quantitative Investing}



\icmlsetsymbol{equal}{*}

\begin{icmlauthorlist}
\icmlauthor{Lakshay Chauhan}{euc}
\icmlauthor{John Alberg}{euc}
\icmlauthor{Zachary C. Lipton}{cmu,amzn}
\end{icmlauthorlist}

\icmlaffiliation{euc}{Euclidean Technologies, Seattle, USA}
\icmlaffiliation{cmu}{Carnegie Mellon University, Pittsburgh, USA}
\icmlaffiliation{amzn}{Amazon AI, Seattle, USA}

\icmlcorrespondingauthor{Lakshay Chauhan}{lakshay.chauhan@euclidean.com}
\icmlcorrespondingauthor{John Alberg}{john.alberg@euclidean.com}
\icmlcorrespondingauthor{Zachary Lipton}{zlipton@cmu.edu}

\icmlkeywords{Machine Learning, ICML}

\vskip 0.3in
]



\printAffiliationsAndNotice{}  

\begin{abstract}
On a periodic basis, publicly traded companies report \emph{fundamentals}, 
financial data including revenue, earnings, debt, among others.
Quantitative finance research has identified several
\emph{factors}, functions of the reported data
that historically correlate with stock market performance.
In this paper, we first show through simulation 
that if we could select stocks via factors 
calculated on future fundamentals (via oracle), 
that our portfolios would far outperform
standard factor models.
Motivated by this insight,
we train deep nets to forecast future fundamentals 
from a trailing 5-year history.
We propose \emph{lookahead factor models} 
which plug these predicted future fundamentals 
into traditional factors.
Finally, we incorporate uncertainty estimates 
from both neural heteroscedastic regression 
and a dropout-based heuristic,
improving performance by adjusting our portfolios to avert risk. 
In retrospective analysis, we leverage
an industry-grade portfolio simulator (backtester)
to show simultaneous improvement in annualized return and Sharpe ratio.
Specifically, the simulated annualized return 
for the uncertainty-aware model 
is 17.7\% (vs 14.0\% for a standard factor model) 
and the Sharpe ratio is 0.84 (vs 0.52).

\end{abstract}

\section{Introduction}
\label{sec:intro}
Public stock markets provide a venue for buying and selling shares, 
which represent fractional ownership of individual companies. 
Prices fluctuate frequently, with the drivers of price movement 
occurring on multiple time scales.  
In the short run, price movements might reflect 
the dynamics of order execution \citep{orderexec1993, bessembinder_2003}  
and the behavior of high frequency traders \citep{hft2016, McGroarty2018}.
On the scale of days, price fluctuation might be driven by the news cycle 
\citep{NBERw18725, oncology2011, schumaker18, CRUZ2013}, 
reports of sales numbers, or product launches \citep{Koku1997}.
In the long run, we expect a company's market value 
to reflect its financial performance as captured in \emph{fundamental data}, 
i.e., reported financial information such as income, revenue, 
assets, dividends, and debt \citep{valuation, Dimson15, mckinsey}.
One popular strategy called \emph{value investing} is predicated
on the idea that the best features for predicting the long-term returns 
on shares in a company are the currently-available fundamental data.

In a typical quantitative (systematic) investing strategy, 
we sort the set of available stocks according to some \emph{factor} 
and construct investment portfolios comprised of those stocks 
which score highest \citep{Dimson15}.
Many quantitative investors engineer \emph{value factors},
typically a ratio of some fundamental to the stock’s price.
Examples include \emph{book-to-market} (the ratio of book value to market value)
and $EBIT/EV$ (earnings before interest and taxes normalized by enterprise value). 
Stocks with high value factor ratios are called \emph{value} stocks 
and those with low ratios are called \emph{growth} 
stocks---presumably the high prices of these stocks 
is predicated on anticipated growth in the future.
A basic premise of value investors is that the market tends 
to systematically over-value growth stocks.
Academic researchers have demonstrated (empirically)
that portfolios that upweight value stocks 
have historically outperformed portfolios that upweight growth stocks 
over the long run \citep{Leivo2017value, fama1992returns}.  

\begin{figure*}[ht!]
    \centering
    \includegraphics[scale=0.3]{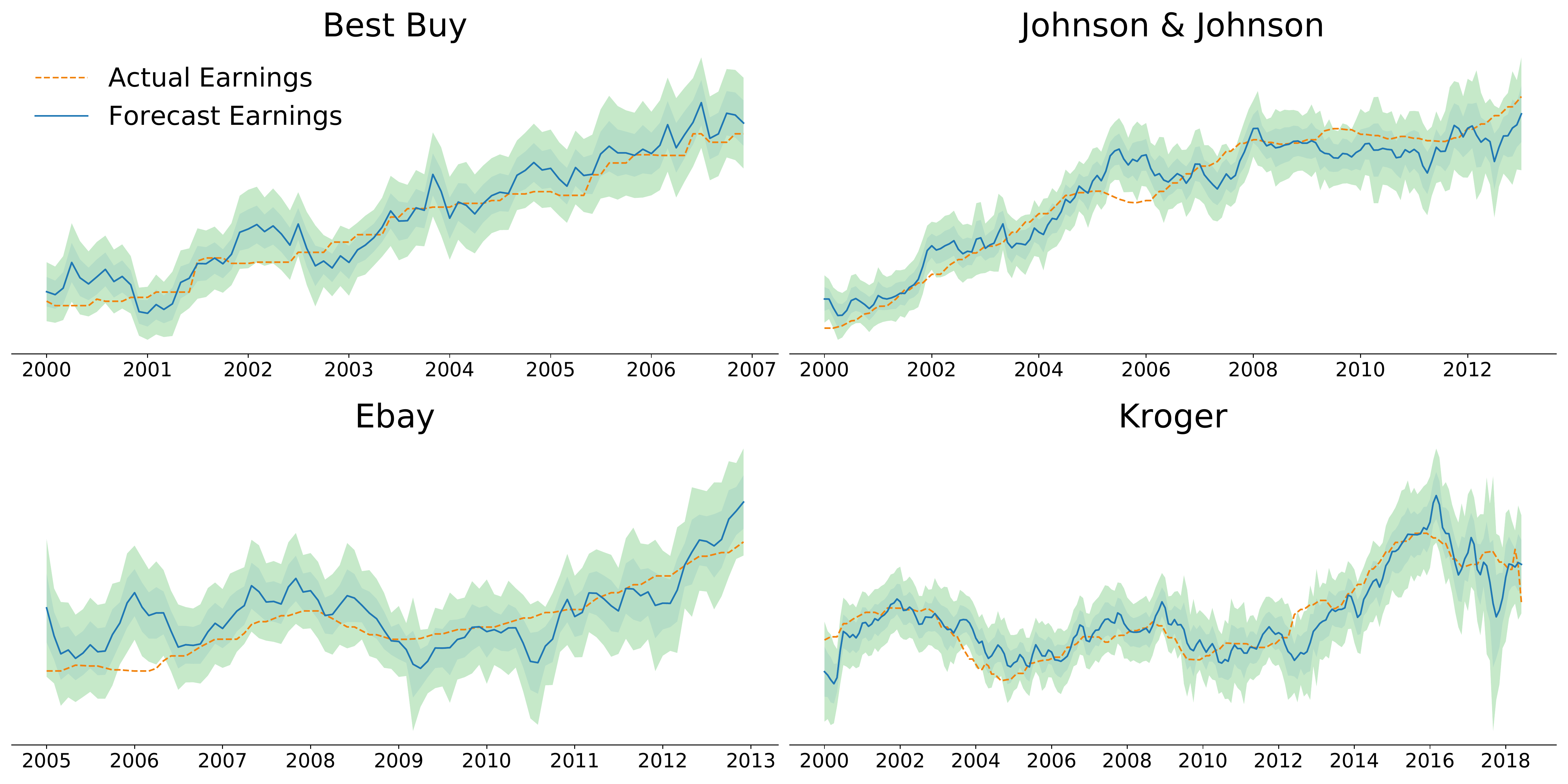}
    \caption{\small{Actual earnings (orange) plotted against deep LSTM forecasts (blue) 
    and uncertainty bounds for selected public companies Best Buy, 
    Johnson \& Johnson, Ebay, and Kroger over different time periods. 
    The LSTM was trained on data from Jan 1, 1970 to Jan 1, 2000. 
    }}
    \label{fig:earnings-plots}
\end{figure*}

\begin{figure}
\centering
\includegraphics[width=.95\linewidth]{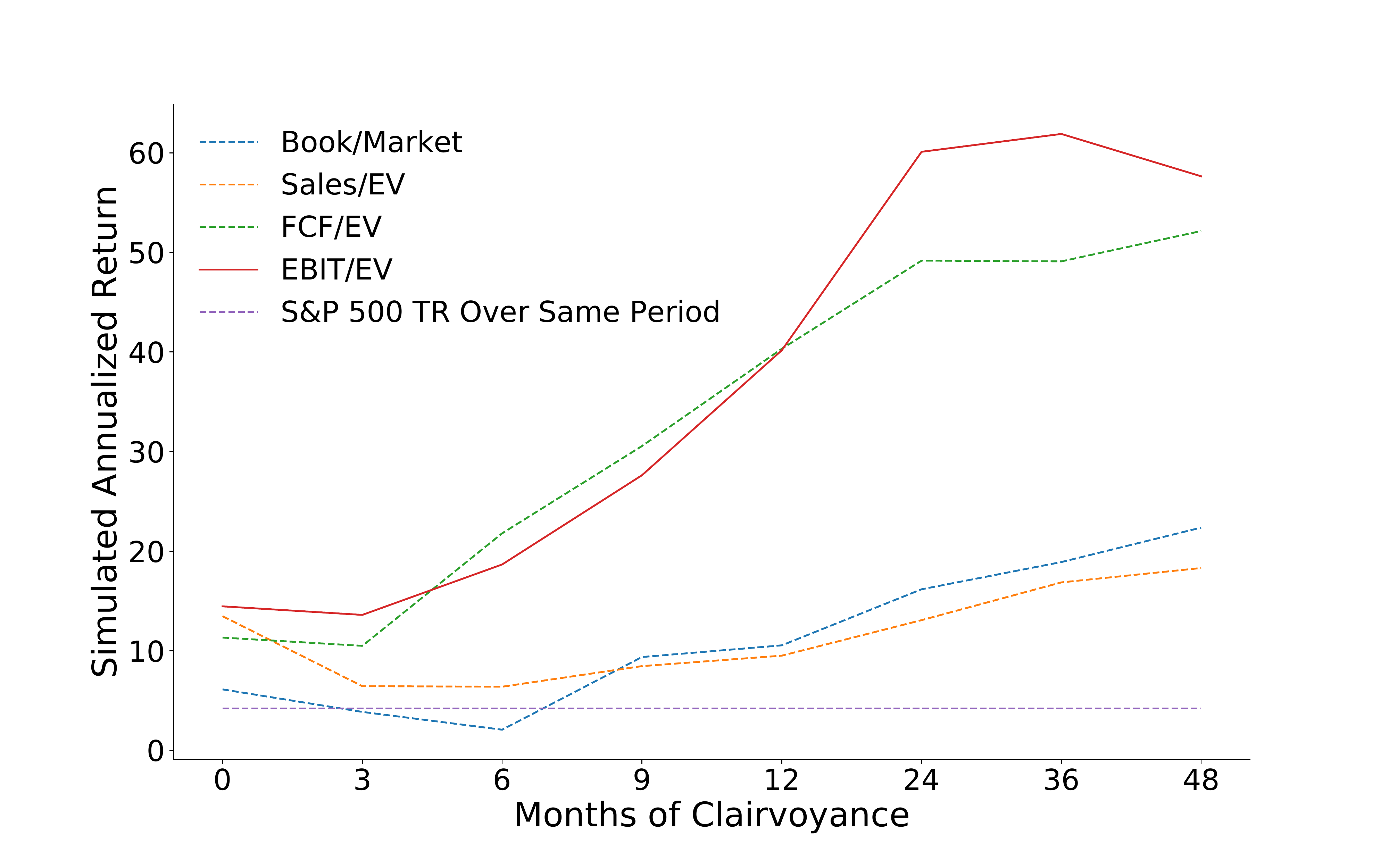}
\caption{Annualized return for various factor models for different degrees of clairvoyance.}
\label{fig:clairvoyant-factor-models}
\vspace{-5px}
\end{figure}

In this paper, we propose an investment strategy 
that constructs portfolios of stocks
based on \emph{predicted future fundamentals}.
Rather than comparing current fundamentals to the current price,
our approach is based on the intuition that 
the long-term success of an investment 
depends on how well the stock is \emph{currently} priced 
relative to \emph{future fundamentals}.
To verify this hypothesis, we run backtests 
with a \emph{clairvoyant model} (oracle) 
that can access future financial reports.

Our experiments show that from Jan 1, 2000 to Dec 31, 2019,
a clairvoyant $EBIT/EV$ factor model 
that perfectly forecasts future $EBIT$ (12 months out) 
would have achieved a $40$\% compound annualized return 
(Figure \ref{fig:clairvoyant-factor-models}). 
That compares to a $14.0\%$ annualized return over the same period 
for a standard factor model using current instead of future earnings. 
While future $EBIT$ (or earnings) are unknowable, 
we hypothesize that some of these gains might be realized
by plugging in forecasted future earnings. 
To test this hypothesis, we investigate the use of deep neural networks (DNNs), 
specifically Multi-Layer Perceptrons (MLPs) and 
Long Short-Term Memory (LSTM) \citep{hochreiter1997long} 
Recurrent Neural Networks (RNNs) to \emph{predict} future earnings 
based on trailing time series of 5 years of fundamental data. 
We denote these models as \emph{Lookahead Factor Models (LFMs)}.

Forecasting future fundamentals is a difficult task
complicated by substantial uncertainty
over both model parameters and inherent noise.
Notably, the problem exhibits heteroscedasticity, 
with noise levels varying across companies and time periods.
For example, a mature consumer goods company 
may have less uncertainty in its future earnings 
than a rapidly growing technology company. 
Moreover, the predictability of earnings might vary across time periods.
For example we expect the task to be more difficult 
in the wake of the 2008 financial crisis than in a comparatively stable period.
In Figure \ref{fig:earnings-plots}, we plot the forecast distribution of future 
earnings against actual earnings for several companies in different time periods.

Classical portfolio allocation theory dictates
that both the expectation and variance of returns
are essential for making decisions.
Therefore, in addition to training DNNs
to generate point forecasts
to be plugged into our factor models,
we also consider methods to forecast the variance 
of the predicted future earnings.
Our uncertainty estimates are derived both via 
neural heteroscedastic regression \cite{ng2017predicting} 
and a popular dropout-based heuristic \citep{Gal2016Uncertainty}.
To construct our uncertainty-aware factor, 
we scale the earnings forecast in inverse proportion 
to the modeled earnings variance.
We show that investment portfolios constructed using this factor exhibit less volatility and enjoy higher returns. 
Simulations demonstrate that investing with LFMs 
based on the risk-adjusted forecast earnings 
achieves a Compound Annualized Return (CAR) of $17.7$\%, 
vs $14.0$\% for a standard factor model and a Sharpe ratio $.84$ vs $.52$.

\section{Related Work}
Deep neural network models have proven useful
for a diverse set of sequence learning tasks,
including machine translation \citep{sutskever2014sequence,bahdanau2014neural}, 
video captioning \citep{mao2014deep,vinyals2015show}, 
video recognition \citep{donahue2015long, tripathi2016context}, 
and time series modeling \citep{lipton2016learning,lipton2016directly, che2016recurrent}.

\paragraph{Deep networks for stock market forecasting}
A number of recent papers consider deep learning approaches 
for predicting stock market performance. 
\citet{batres2015deep} use MLPs for stock market prediction 
and \citet{qiu16} use neural networks
to predict the direction of a broad market index 
using technical indicators such as price.
\citet{ding2015deep} use recursive tensor nets 
to extract events from CNN news reports
and use convolutional neural nets 
to predict future performance 
from a sequence of extracted events. 
Several researchers have considered deep learning 
for stock-related predictions \citep{chen2015lstm,wanjawa2014ann,jia2016investigation},
however, in all cases, the empirical studies 
are limited to few stocks or short time periods.
To our knowledge, an early version of this paper 
was the first public work to apply modern deep networks 
to large-scale time series of financial fundamentals data 
and the first to introduce lookahead factor models.

\paragraph{Uncertainty Estimation}
Approaches for estimating uncertainty are differentiated
both according to what notion of uncertainty they address
and by the methods they employ.
For example, some methods address uncertainty in the model parameters
\citep{Gal2015DropoutC, Gal2015Dropout, Gal2017Concrete, Balaji2017, Heskes97practicalconfidence, NixWeig1994}
while others use neural networks to directly output prediction intervals 
\citep{pmlr-v80-pearce18a, EIM2018, khosravi2011}. 
Prediction uncertainty can be decoupled 
into \emph{model uncertainty} (epistemic uncertainty) 
and the inherent noise due to conditional variability 
in the label (aleatoric uncertainty) \citep{der2009aleatory, kendall2017uncertainties}.
Epistemic uncertainty can arise from uncertainty 
over the value of the model parameters 
and/or model structure. 
In contrast, aleatoric uncertainty owes to the inherently stochastic nature of the data.
Note that yet other sources exist 
but are unaccounted for in this dichotomy,
e.g., uncertainty due to distribution shift. 

%
%
%
%
Assuming heteroscedasticity, i.e., that the noise is data dependent,
\citet{NixWeig1994, Heskes97practicalconfidence} 
train two neural networks, one to estimate the target value 
and another to estimate the predictive variance. 
More recently, \citet{ng2017predicting} used a single network
to forecast both the mean and the conditional variance
when predicting surgery durations.
\citet{NixWeig1994, Balaji2017} 
use the bootstrap method where multiple networks are trained 
on random subsets of the data
to obtain uncertainty estimates.

Bayesian Neural Networks (BNN),
learn an approximate posterior distribution over model parameters,
enabling the derivation of predictive distributions
and thus estimates of epistemic uncertainty.
In one approach to training BNNs,
\citet{blundell2015weight} employs variational inference, 
choosing a variational distribution 
consisting of independent Gaussians---thus 
each weight is characterized by two parameters.
Then, employing the reparamterization trick, 
they optimize the variational parameters 
by gradient descent in a scheme they call \emph{Bayes-by-backprop}.
In this paper, we follow the related work of 
\citet{Gal2015DropoutC, Gal2017Concrete, Gal2016Uncertainty},
who propose Monte Carlo dropout (MC-dropout),
a heuristic that obtains uncertainty estimates by using dropout during prediction. 
Their approach is based on insights from analysis 
establishing a correspondence between stochastically-regularized 
neural networks and deep Gaussian processes.

\section{Data}
\label{sec:data}
In this research, we consider all stocks 
that were publicly traded on the NYSE, NASDAQ, or AMEX exchanges 
for at least $12$ consecutive months 
between Jan 1, 1970, and Dec 31, 2019.
From this list, we exclude non-US-based companies,
financial sector companies, and any company 
with an inflation-adjusted market capitalization value below  
$100$ million dollars in January 1, 2010 terms.
The final list contains $12,415$ stocks. 
Our features consist of reported financial information 
as archived by the \emph{Compustat North America} and \emph{Compustat Snapshot} databases. 
Because reported information arrives intermittently throughout a financial period,
we discretize the raw data to a monthly time step.
Because we are interested in long-term predictions
and in order to smooth out seasonality in the data,
at every month, we feed a time-series of inputs 
with a one year gap between time steps and predict the earnings 
one year into the future from the last time step. 
For example, one trajectory in our dataset
might consist of data for a given company from
(Jan 2000, Jan 2001, ..., Jan 2007),
which are used to forecasts earnings for Jan 2008. 
For the same company, we will also have another trajectory 
consisting of data from (Feb 2000, Feb 2001, ..., Feb 2007)
that are used to forecast earnings for Feb 2008.
Although smaller forecast periods such as 3 or 6 months may be easier to forecast,
we use a forecast period of 12 months as it provides 
the best trade-off between model accuracy and portfolio performance. 
We discuss this trade-off in Figure \ref{fig:fcst_period} in Section \ref{sec:experiments}.

Three classes of time-series data are used at each time-step $t$: 
fundamental features, momentum features, and auxiliary features.
For fundamental features, income statement and cash flow items 
are cumulative \emph{trailing twelve months}, denoted TTM, 
and balance sheet items are of the \emph{most recent quarter}, denoted MRQ.
TTM items include revenue; 
cost of goods sold;
selling, general \& admin expense;
earnings before interest and taxes or $EBIT$;
and free cash flow, defined as operating cash flow minus capital expenditures.
MRQ items include
cash and cash equivalents; 
receivables; 
inventories; 
other current assets; 
property plant and equipment; 
other assets; 
debt in current liabilities; 
accounts payable; 
taxes payable; 
other current liabilities; 
shareholders' equity; 
total assets; 
and total liabilities.
For all features, we deal with missing values 
by filling forward previously observed values,
following the methods of \citet{lipton2016learning}.
Additionally, we incorporate $4$ \emph{momentum features},
which indicate the price movement of the stock 
over the previous $1$, $3$, $6$, and $9$ months, respectively. 
So that our model picks up on relative changes
and does not focus overly on trends in specific time periods,
we use the percentile among all stocks as a feature (vs absolute numbers).

Finally, we consider a set of auxiliary features 
that include a company's short interest
(\% of a company's outstanding shares that are held short); 
a company's industry group as defined by Standard and Poor's GICS code 
(encoded as a $27$ element one-hot vector with $26$ industry groups 
plus one for indicating an unknown industry classification);
and the company's size category of micro-cap, small-cap, mid-cap, 
and large-cap (encoded as a one-hot vector). 


There can be wide differences in the absolute value of the 
fundamental features described above when compared between 
companies and across time. 
For example, Exxon Mobil's annual revenue 
for fiscal 2018 was \$279 billion USD
whereas Zoom Video Communications 
had revenue of \$330 million USD for the same period.
Intuitively, these statistics are more meaningful
when scaled by some measure of a company's size. 
In preprocessing, we scale all fundamental features in a given time series 
by the market capitalization in the last input time-step of the series. 
We scale all time steps by the same value so that the DNN 
can assess the relative change in fundamental values between time steps. 
While other notions of size are used, 
such as enterprise value and book equity, 
we choose to avoid these measures 
because they can, although rarely, take negative values. 
We then further scale the features so that they each individually 
have zero mean and unit standard deviation.

\section{Methods}
\label{sec:methods}

\subsection{Forecasting Model}
We divide the timeline into 
\textit{in-sample} and \textit{out-of-sample} periods. 
Data in the in-sample period range from Jan 1, 1970 
to Dec 31, 1999 (1.2M data points),
while out-of-sample test data
range from Jan 1, 2000, to Dec 31, 2019 (1M data points). 
Since we do not want to overfit the finite training sample, 
we hold out a validation set by randomly selecting
30\% of the stocks from the in-sample period. 
We use this in-sample validation set to tune the hyperparameters 
including the initial learning rate, model architecture,
and objective function weights. 
We also use this set to determine early-stopping criteria.
When training, we record the validation set accuracy after each epoch, 
saving the model for each best score achieved. 
We halt the training if the model doesn't improve for 25 epochs 
and select the model with best validation set performance. 
In addition to evaluating how well our model 
generalizes on the in-sample holdout set,
we evaluate whether the model successfully 
forecasts fundamentals in the \textit{out-of-sample} period. 

Financial fundamental data is inherently temporal.
Our methods apply LSTM and MLP models to forecast a company's 
future financial fundamental data from past fundamental data.
We choose each input to consist of a five year window of data
with an annual time step.
Companies with less than five years of historical fundamentals 
are excluded from the training and testing set. 
As output, we are interested in predicting $EBIT$ 
(earnings before interest and taxes) 
twelve months into the future because forecasted $EBIT$ 
is required to compute the factor that drives our investment models.

Previously, we tried to predict relative returns 
directly (using price data) with an LSTM model.
While the LSTM outperformed other approaches on the in-sample data, 
it failed to meaningfully outperform a linear model on the out-of-sample data. 
Given only returns data as targets, the LSTM easily overfit the training 
data while failing to improve performance on in-sample validation.
While the price movement of stocks 
is known to be extremely noisy, 
we suspected that temporal relationships among the fundamental data 
may exhibit a larger signal to noise ratio,
and this intuition motivates us to focus
on forecasting future fundamentals.

Although we only need an estimate of the fundamental feature \textit{EBIT}
in order to construct our factor, 
we forecast all 17 fundamental features. 
One key benefit of our approach is that 
by doing \textit{multi-task learning} \citep{mtl1, mlt_seb}, 
predicting all 17 fundamentals, 
we provide the model with considerable
training signal so that it is less susceptible to overfitting.
We also predict the uncertainty (or risk) for those targets 
(described in detail in the next section).
Since we care more about EBIT over other outputs,
we up-weight it in the loss 
(introducing a hyperparameter $\alpha_{1}$). 
For LSTMs, the prediction at the final time step 
is more important and hence we use hyperparameter $\alpha_{2}$ 
to up-weight the loss for the final time step. 


During experimentation we examined several hyperparameters.
We clip the gradients,
rescaling to a maximum L2 norm to avoid exploding gradients.
We constrain the maximum norm of 
the vector of weights incoming to each hidden unit.
We also experiment with L2 regularization
and dropout for further regularization.
Our MLPs use \textit{ReLu} activation functions 
in all hidden layers.
Our models tend to be sensitive to the 
weight initialization hyperparameters.
Based on validation performance, 
we settled on \textit{GlorotUniform Intialization} 
\citep{Glorot10},
which made results more consistent across runs.
We also use batch normalization \citep{batchnorm}.
Each model is an ensemble of 10 models 
trained with a different random seed.
For LSTM models, in addition 
to the hyperparameters discussed above, 
we use recurrent dropout 
to randomly mask the recurrent units.

We use a genetic algorithm \citep{Goldberg} 
for hyperparameter optimization. 
The optimizer \emph{AdaDelta} \citep{adadelta} is used 
with an initial learning rate of $0.01$. 
It took $150$ epochs to train an ensemble on a machine 
with $16$ Intel Xeon E5 cores and $1$ Nvidia P100 GPU. 
The final hyperparameters as a result of the optimization process 
are presented in Table \ref{tab:hyperparams}.

\begin{table}[ht!]
\centering
\vspace{-15px}
 \caption{\small{MLP, LSTM Hyperparameters}}
 \label{tab:hyperparams}
 \begin{tabular}{l|cc}
  \toprule
  Hyperparameter & \small{MLP} &\small{LSTM}\\
  \midrule
    \small{Batch Size}& $256$ & $256$  \\
    \small{Hidden Units}& $2048$ & $512$  \\
    \small{Hidden Layers}& $1$ & $1$  \\
    \small{Dropout}&  $0.25$ & $0.0$  \\
    \small{Recurrent Dropout}& n/a & $0.25$  \\
    \small{Max Gradient Norm}& $1$ & $1$  \\
    \small{Max Norm} & $3$ & $3$ \\
   {$\alpha_1$ }& $0.75$ & $0.5$  \\
   {$\alpha_2$ }& n/a & $0.7$  \\
  \bottomrule
\end{tabular}    
\end{table}

\subsection{Uncertainty Quantification}

We model the targets as conditionally Gaussian distributed
about a mean $f_{\theta}(\mathbf{x})$,
which is the predicted output.
While standard least squares regression relies on the assumption
that the noise level is independent of the inputs $\mathbf{x}$,
we jointly model the conditional mean and variance,
denoting our model variance by $g_{\theta}(\mathbf{x})$.
Following \citet{ng2017predicting}, we model this heteroscedasticity
by emitting two outputs for each target variable in the final layer:
one set of outputs corresponds to the predicted means 
of the target values $f_{\theta}(\mathbf{x})$
and the second half predicts the variance 
of the output values $g_{\theta}(x)$. 
We use the \emph{softplus} activation function 
for outputs corresponding to the variance $g_{\theta}(x)$ 
to ensure non-negativity.
The predictors share representations 
(and thus parameters for all representation layers)
and are differentiated only at the output layer.
To learn the network parameters $\theta$,
we train the neural network with the maximum likelihood objective as follows:

\begin{align*}
\boldsymbol{\theta}^{\text{MLE}}
= &\max_{\theta} \prod_{i=1}^{n} \frac{1}{\sqrt{2\pi g_{\theta}(\mathbf{x}_i)^2}} \exp \left( \frac{-(y_i-f_{\theta}(\mathbf{x_i}))^2}{2 g_{\theta}(\mathbf{x}_i)^2} \right)\\
= & \min_{\boldsymbol{\theta}} \sum_{i=1}^{n}
\left( \log(g_{\theta}(\mathbf{x}_i)) + \frac{(y_i-f_{\theta}(\mathbf{x_i}))^2}{2 g_{\theta}(\mathbf{x}_i)^2}  \right).
\end{align*}



In the above loss function, the first term 
penalizes large uncertainty in the model. 
This allows the DNN to minimize 
the prediction interval width and provide meaningful bounds. 
The second term penalizes an over-confident model (low uncertainty) 
with high error focusing on model accuracy. 

To estimate the epistemic uncertainty,
we train the DNN model using dropout 
and leverage a heuristic by \citet{Gal2015DropoutC}
that applies dropout during prediction. 
Model variance is given by the variance in the outputs
across $10$ Monte Carlo draws of the dropout mask 
where the dropout rate is $0.25$.
The number $10$ is selected based on the maximum number of parallel executions
that could be launched on the computing infrastructure.
Hence the total variance is given by the sum of model variance 
(variance in the predictions) and noise variance (predicted variance).
%
%
%
In summary, the final model is an ensemble of 10 equally-weighed DNN models 
with different random seeds for dropout. 
Variance is estimated as a sum of the variance across dropout forward passes 
and the estimated input-conditioned noise $g_{\theta}(\mathbf{x})$.

\subsection{Quantitative Factor Models}
Typical quantitative investment strategies use factors such as $EBIT/EV$ 
to construct portfolios by investing in the stocks that score highest. 
Whereas a standard Quantitative Factor Model (QFM) uses current $EBIT$, 
we are interested in comparing such investment strategies 
with strategies that use forecast $EBIT$.
We construct a look-ahead factor $EBIT_m/EV$ for each model $m$,
where $EBIT_m$ is the model's \emph{forecast} $EBIT$.
Hence there is a LFM for auto-regression (LFM Auto Reg), 
multivariate linear model point forecast (LFM Linear), 
multi-layer perceptron point forecast (LFM MLP), 
LSTM point forecast (LFM LSTM),
variance scaled MLP forecast (LFM UQ-MLP),
and variance scaled LSTM forecast (LFM UQ-LSTM).

Variance scaled models (UQ-MLP, UQ-LSTM) incorporate
uncertainty over the forecasted $EBIT$
to reduce the risk of the portfolio.
Two companies with the same $EBIT$ might have very different levels of uncertainty.
The one with higher uncertainty around $EBIT$ 
(higher variance) is more risky for investors.
Such a company will not only increase the portfolio risk
but also decrease the expected returns due to higher forecast error.
Hence, we scale the $EBIT$ in inverse proportion 
to the total variance for the LFM UQ models.
A portfolio created with the \textit{risk-adjusted}
look-ahead factor $\frac{EBIT}{\sigma^2 EV}$
is expected to have lower average volatility of earnings
than a portfolio created using the $\frac{EBIT}{EV}$ factor.


\section{Portfolio Simulation}
\label{sec:simulation}
While we train and evaluate our models using the negative log likelihood objective,
for our purposes, this metric is merely a surrogate measure of performance. 
What investors actually care about is a portfolio's performance 
in terms of both return and risk (volatility) over some specified time period. 
To establish a correspondence between 
predictive performance and investment returns,
we employ an industry-grade investment simulator.

The goal of the simulator is to recreate as accurately as possible 
the investment returns an investor would have achieved 
had they been using the model over a specific period of time 
and within a specific universe of stocks. 
To this end, the simulation must incorporate transaction costs,
liquidity constraints, bid-ask spreads, 
and other types of friction that exist in the management 
of a real-life portfolio of stocks. 

\begin{figure}[t!]
    \centering
    \includegraphics[scale=0.23]{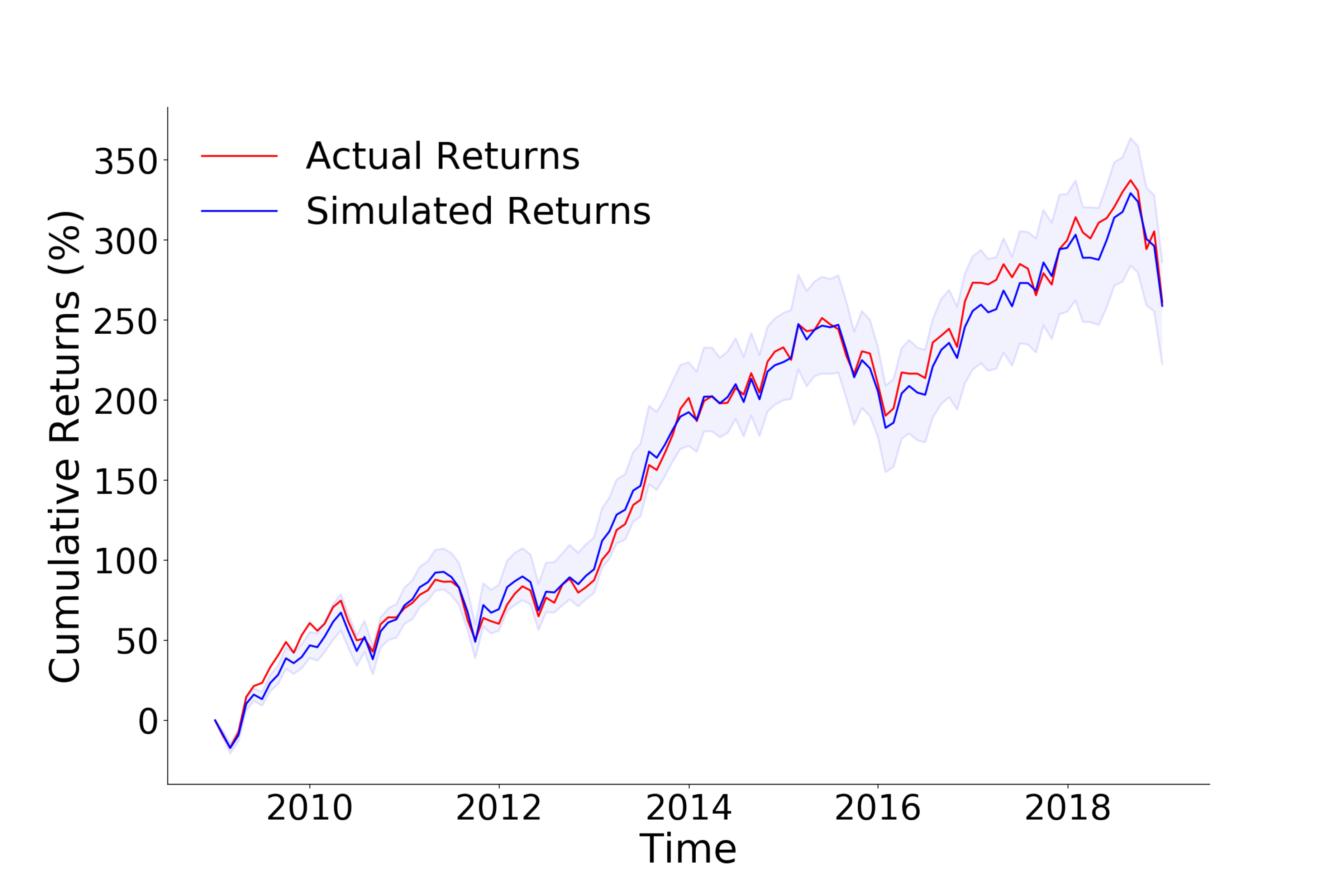}
    \caption{\small{Simulated returns of a quantitative (systematic) strategy vs. 
    the returns generated from live trading of the same strategy 
    over the period Jan 1, 2009 to Dec 31, 2018. 
    The shaded band represents the distribution of simulated returns 
    generated by varying the initial state of the portfolio at $t_0$. 
    }}
    \label{fig:sim_acc}
\end{figure}

The simulation algorithm works as follows:
We construct portfolios by ranking all stocks 
according to the factor of interest
and invest equal amounts of capital into the top $50$ stocks,
re-balancing monthly.
We limit the number of shares of a security bought or sold in a month
to no more than $10\%$ of the monthly volume for a security. 
Simulated prices for stock purchases and sales are based 
on the volume-weighted daily closing price of the security 
during the first $10$ trading days of each month.
If a stock paid a dividend during the period it was held, 
the dividend was credited to the simulated fund 
in proportion to the shares held.
Transaction costs are factored in as $\$0.01$ per share, 
plus an additional slippage factor 
that increases as a square of the simulation’s volume participation in a security. 
Specifically, if participating at the maximum $10\%$ of monthly volume, 
the simulation buys at $1\%$ more than the average market price 
and sells at $1\%$ less than the average market price. 
This form of slippage is common in portfolio simulations 
as a way of modeling the fact that as an investor's 
volume participation increases in a stock, 
it has a negative impact on the price of the stock for the investor. 

Monthly return values $r_t$ are determined 
by calculating the percentage change in total portfolio value 
between the beginning and end of each simulated month. 
From the sequence of monthly portfolio return values 
and knowledge of the annualized risk free 
rates of return $R^f$ over the same period, 
we compute standard portfolio performance statistics 
such as the Compound Annualized Return ($\textit{CAR}$) and the Sharpe ratio. 
These are defined as follows:
\begin{gather}
    \textit{CAR} = \Big[\prod_{t}^{T} (r_t+1)\Big]^{12/T} - 1 \\
    \textit{Sharpe Ratio} = \frac{\textit{CAR} - R^f}{\sqrt{12}\sigma},
\end{gather}
where $\sigma$ is the standard deviation of monthly portfolio returns $r_t$. 
The Sharpe ratio is commonly used as a measure of risk adjusted portfolio performance.

Due to how a portfolio is initially constructed and the timing of cash flows, 
two portfolio managers can get different investment results 
over the same period using the same quantitative model. 
To account for this variation, we run $300$ portfolio simulations for each model 
where each portfolio is initialized from a randomly chosen starting state. 
The portfolio statistics, such as $\textit{CAR}$ and Sharpe ratio, 
that are presented in this paper are 
the mean of statistics generated by the $300$ simulations.

To illustrate the accuracy of the simulator,
we compare the returns 
generated by the simulator with the actual returns 
generated by a quantitative strategy 
that was executed between Jan 1, 2009 and Dec 31, 2018
(Figure \ref{fig:sim_acc}). 
In this case, it can be clearly seen that 
the distribution of returns generated by the $300$ portfolios 
with different random initial starting states almost always 
encompassed the actual returns of the quantitative strategy. 
Furthermore, the mid-point of the simulated returns 
tracks the actual returns very closely.

\section{Experiments}
\label{sec:experiments}

\begin{figure}[ht!]
    \centering
    \includegraphics[scale=0.18]{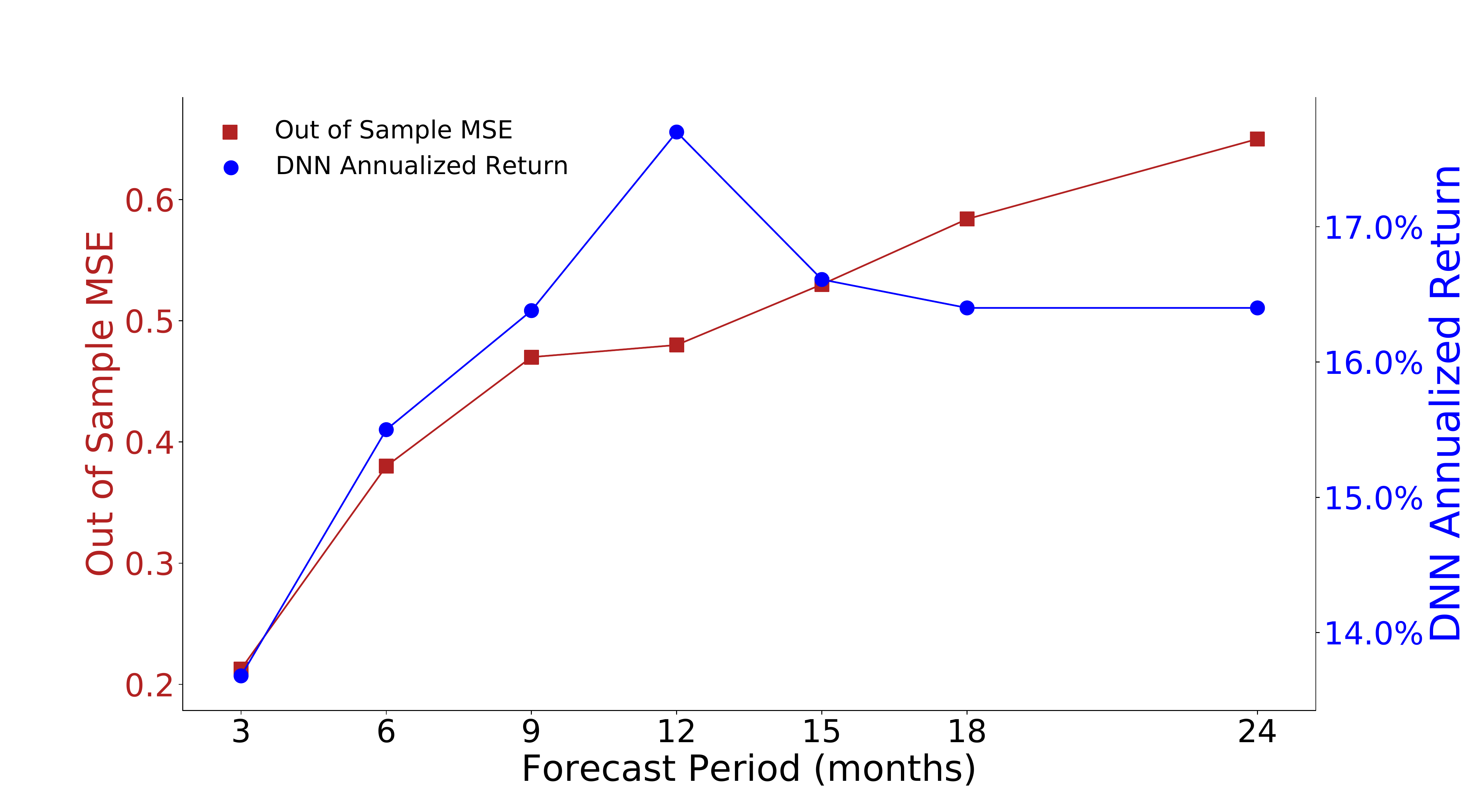}
    \caption{\small{ 
    MSE (red) of the out-of-sample period
    2000-2019 increases with forecast period length. 
    The forecasting model becomes less accurate the further we go out in the future.
    Simulated returns of the LSTM model (blue) increase with
    forecast period up to 12 months and then start decreasing.
    }}
    \label{fig:fcst_period}
\end{figure}

Recall that in Section \ref{sec:intro},
we demonstrated that if we could forecast $EBIT$ perfectly
(the \textit{clairvoyant model}), 
the portfolios built using the lookahead factor 
would far outperform standard factor models.
Of course, perfect knowledge of future $EBIT$ is impossible,
but we speculate that by forecasting future $EBIT$
we can also realize some of these gains,
outperforming standard factor models. 
The question arises as to how far into the future to forecast.
Clearly forecasting becomes more difficult
the further into the future we set the target.
In Figure \ref{fig:fcst_period}, 
we plot the out-of-sample MSE for different forecast periods.
The further we try to predict,
the less accurate our model becomes.
However, at the same time, the clairvoyance study 
(Figure \ref{fig:clairvoyant-factor-models}) tells us 
that the value of a forecast increases monotonically
as we see further into the future. 
In our experiments, the best trade-off is achieved
with a forecasting period of 12 months as shown 
by the blue curve in Figure \ref{fig:fcst_period}. 
Simulated returns increase as the forecasting window lengthens
up until 12 months after which the returns start to fall. 

Motivated by our study with clairvoyant factor models,
we first establish a correspondence between 
the accuracy of DNN forecasts and portfolio returns. 
While training the LSTM model, we checkpoint 
our model's parameters after each epoch. 
These models have sequentially decreasing mean squared error. 
Once training is complete, for each saved model 
we generate $EBIT$ forecasts for the out-of-sample period.
We then use the forecasts to generate corresponding portfolio returns 
by simulating the portfolios constructed using the forecast $EBIT/EV$ factors.
As a result, we have a sequence of $MSE$ and portfolio return pairs,
allowing us to evaluate the correspondence 
between decreasing $MSE$ and portfolio return.

\begin{figure}[h!]
    \centering
    \includegraphics[scale=0.19]{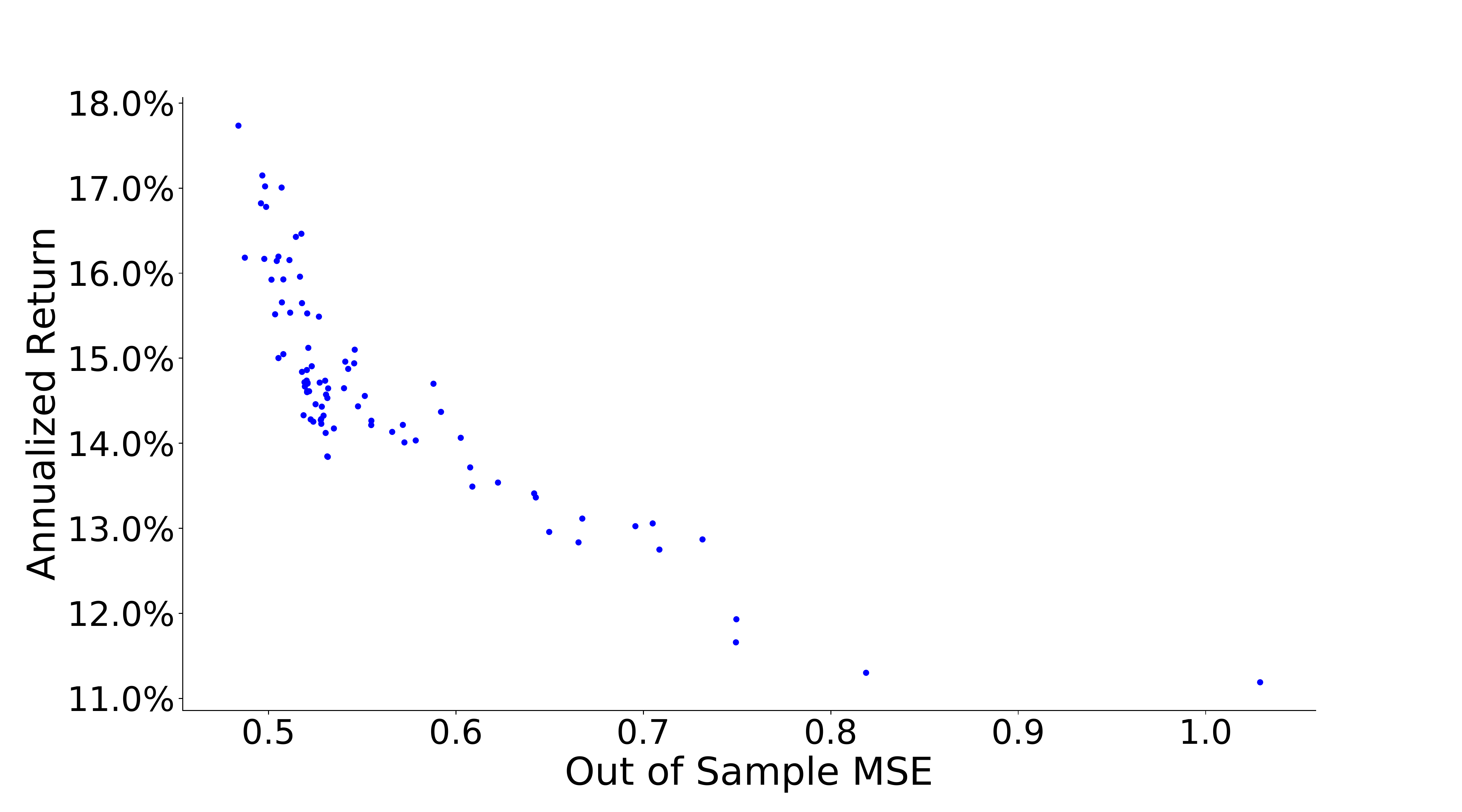}
    \caption{\small{Correspondence between the LSTM model accuracy and portfolio returns. 
    Bottom-rightmost point is evaluated after the first epoch. 
    As training progresses, points in the graph move towards the upper left corner. 
    Portfolio returns increase as the model accuracy improves 
    (out-of-sample MSE decreases).
    }}
    \label{fig:acc_ret}
\end{figure}

Figure \ref{fig:acc_ret} shows the relationship 
between increasing model accuracy and improving portfolio returns. 
This experiment validates our hypothesis that returns 
are strongly dependent on the accuracy of the forecasting model.

\begin{figure}[h!]
    \centering
    \includegraphics[scale=0.23]{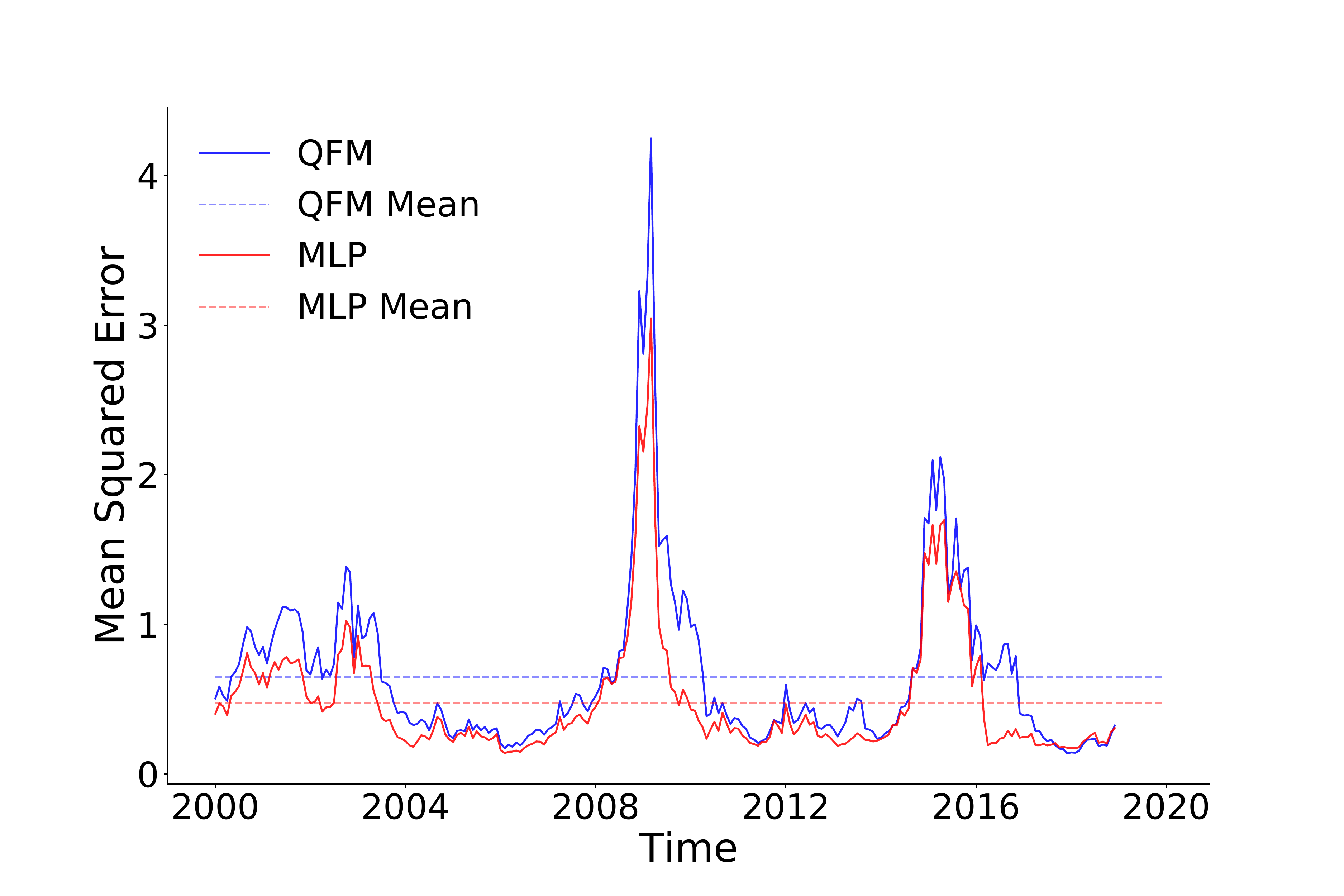}
    \caption{\small{MSE over out-of-sample time period for MLP (red)
    and QFM or Naive predictor (blue)
    }}
    \label{fig:mse_time}
\end{figure}

\begin{figure*}[t!]
    \centering
    \vspace{-5px}
    \includegraphics[scale=0.30]{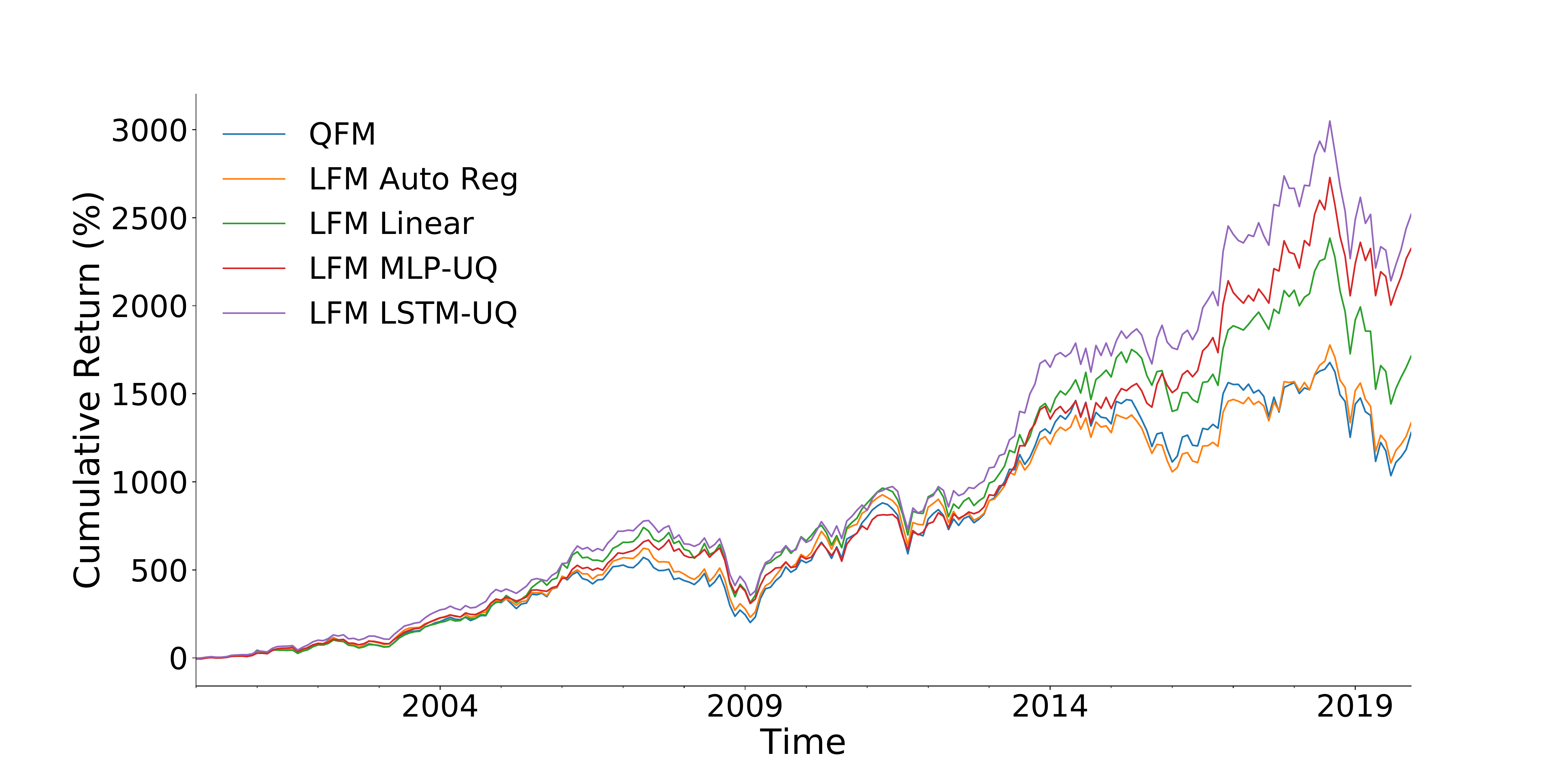}
    \vspace{-10px}
    \caption{Cumulative return of different strategies  
    for the out-of-sample period.
    LFM UQ-LSTM consistently outperforms throughout the entire period.
    }
    \label{fig:total_return}
\end{figure*}

As a first step in evaluating the forecast produced by the neural networks, 
we compare the MSE of the predicted fundamentals on out-of-sample data 
with a naive predictor where the predicted fundamentals at time \textit{t} 
are assumed to be the same as the fundamentals
at \textit{t -- 12}. In nearly all the months,
however turbulent the market, neural networks outperform 
the naive predictor (Figure \ref{fig:mse_time}).

\begin{table}[ht!]
    \centering
    \caption{\small{Out-of-sample performance for the 2000-2019 time period.
    All factor models use EBIT/EV. QFM uses current EBIT while our proposed 
    LFMs use predicted future EBIT.}}
    \vspace{10px}
    \begin{tabular}{l|ccc}
    \toprule
    Strategy & MSE & CAR & Sharpe Ratio \\
    \midrule
        \small{S\&P 500 }& n/a & $6.05\%$ & $0.32$ \\
        \small{QFM} & $0.65$ & $14.0\%$ & $0.52$ \\
        \small{LFM Auto Reg} & $0.58$ & $14.2\%$ & $0.56$ \\
        \small{LFM Linear} & $0.52$ & $15.5\%$ & $0.64$ \\
        \small{LFM MLP} & $0.48$ & $16.1\%$ & $0.68$ \\
        \small{LFM LSTM} & $0.48$ & $16.2\%$ & $0.68$ \\
        \small{LFM UQ-LSTM} & $0.48$ & $\mathbf{17.7\%}$ & $\mathbf{0.84}$ \\
        \small{LFM UQ-MLP} & $0.47$ & $\mathbf{17.3\%}$ & $\mathbf{0.83}$ \\
    \bottomrule
    \end{tabular}
    \label{tab:mse_portfolio_perf}
\end{table}

Table \ref{tab:mse_portfolio_perf} demonstrates a clear advantage 
of using look-ahead factor models or LFMs over standard QFM. 
MLP and LSTM LFMs achieve higher model accuracy 
than linear or auto-regression models and thus 
yield better portfolio performance. 
Figure \ref{fig:total_return} shows the cumulative return 
of all portfolios across the out-of-sample period.

Investors not only care about return of a portfolio
but also the risk undertaken as measured by volatility. 
Risk adjusted return or Sharpe ratio is meaningfully higher for LFM UQ models 
which reduce the risk by scaling the $EBIT$ forecast 
in inverse proportion to the total variance.

\begin{table}[H]
    \centering
    \caption{\small{Pairwise t-statistic for Sharpe ratio. 
    The models are organized in increasing order of Sharpe ratio values.
    t-statistic for LFM UQ models are marked in bold 
    if they are significant with a significance level of $0.05$. 
    }}
    \vspace{10px}
    \begin{tabular}{l|cccccc}
    \toprule
     & \tiny{Auto-Reg} & \tiny{Linear} & \tiny{MLP} & \tiny{LSTM} & \tiny{UQ-LSTM} & \tiny{UQ-MLP} \\
    \midrule
        \tiny{QFM} & \tiny{$0.76$} & \tiny{$2.52$} & \tiny{$2.93$} & \tiny{$2.96$} & \tiny{$\mathbf{5.57}$} & \tiny{$\mathbf{6.01}$} \\
        \tiny{Auto Reg} & & \tiny{$1.89$} & \tiny{$2.31$} & \tiny{$2.36$} & \tiny{$\mathbf{5.10}$} & \tiny{$\mathbf{5.57}$} \\ 
        \tiny{Linear} & & & \tiny{$0.36$} & \tiny{$0.46$} & \tiny{$\mathbf{3.12}$} & \tiny{$\mathbf{3.66}$} \\
        \tiny{MLP} & & & & \tiny{$0.10$} & \tiny{$\mathbf{2.82}$} & \tiny{$\mathbf{3.39}$} \\
        \tiny{LSTM} & & & & & \tiny{$\mathbf{2.66}$} & \tiny{$\mathbf{3.22}$} \\
    \bottomrule
        
    \end{tabular}
    \label{tab:t_stat_sharpe}
\end{table}

\begin{table}[!h]
    \centering
    \caption{Cross section of monthly returns. 
    The universe of stocks is ranked by the given factor and
    divided into 10 groups of equally weighted stocks.
    The top decile (marked as \textit{High}) is formed 
    by the top 10\% of the stocks ranked by the factor 
    and the bottom decile (marked as \textit{Low}) 
    is formed by the bottom 10\% of the rankings. 
    \textit{H -- L} represents the factor
    premium.}
    \vspace{10px}
    \begin{tabular}{r|ccc}
    \toprule
        \small{Decile} & \small{QFM} & \small{LSTM} & \small{UQ-LSTM} \\
    \midrule
        \small{H\tiny{igh}} \small{1} & \small{$1.39$} & \small{$1.38$} & \small{$1.47$} \\
        \small{2} & \small{$1.24$} & \small{$1.21$} & \small{$1.18$} \\
        \small{3} & \small{$1.15$} & \small{$1.12$} & \small{$1.13$} \\
        \small{4} & \small{$1.16$} & \small{$1.08$} & \small{$1.04$} \\
        \small{5} & \small{$1.06$} & \small{$1.14$} & \small{$0.97$} \\
        \small{6} & \small{$1.00$} & \small{$1.04$} & \small{$0.98$} \\
        \small{7} & \small{$0.95$} & \small{$0.94$} & \small{$0.98$} \\
        \small{8} & \small{$0.85$} & \small{$0.75$} & \small{$0.90$} \\
        \small{9} & \small{$0.78$} & \small{$0.79$} & \small{$0.74$} \\
        \small{L\tiny{ow}} \small{10} & \small{$0.73$} & \small{$0.57$} & \small{$0.64$} \\
    \midrule
        \small{H -- L} & \small{$0.66$} & \small{$0.80$} & \small{$0.83$} \\
    \midrule
        \small{t-statistic} & \small{2.31} & \small{2.78} & \small{3.57} \\
    \bottomrule
    \end{tabular}
    \label{tab:cross_section}
\end{table}

We provide pairwise t-statistics for Sharpe ratio in Table \ref{tab:t_stat_sharpe},
where improvement in Sharpe ratio for LFM UQ models is statistically significant.
As discussed in Section \ref{sec:simulation}, 
we run $300$ simulations with varying initial start state for each model. 
Additionally, we randomly restrict the universe of stocks 
to $70\%$ of the total universe making the significance test 
more robust to different portfolio requirements.
We aggregate the monthly returns of these $300$ simulations 
by taking the mean and perform bootstrap resampling on the monthly returns 
to generate the t-statistic values for Sharpe ratio shown in Table \ref{tab:t_stat_sharpe}. 
The last two columns corresponding to LFM UQ models 
provide strong evidence that the Sharpe ratio is significantly improved 
by using the estimated uncertainty to reduce risk.

In addition to providing simulation results 
of concentrated 50 stock portfolios
(Table \ref{tab:mse_portfolio_perf}), 
we also provide the cross section of returns 
generated for the models LFM-LSTM and LFM UQ-LSTM
on the out-of-sample period (Table \ref{tab:cross_section}). 
The cross section is constructed by sorting stocks by each factor 
and splitting them into $10$ equally sized portfolios 
ranging from the top decile (highest factor values)
to the bottom decile (lowest factor values). 
The portfolios are rebalanced quarterly according to the factor sorts. 
The cross section shows the efficacy of the factor 
when looked at across the entire investment universe, 
where monthly returns increase almost monotonically 
as we go from the bottom decile to the top decile. 
The difference between the top and bottom decile (high minus low or $H-L$)
is called the factor premium. 
The $t$-statistic for the factor premium is significant
and greater for UQ-LSTM than LSTM and QFM (Table \ref{tab:cross_section}).

\section{Conclusion}
\label{sec:discussion}
In this paper, we demonstrate that by predicting fundamental data with deep learning,
we can construct lookahead factor models that 
significantly outperform equity portfolios based on traditional factors.
Moreover, we achieve further gains 
by incorporating uncertainty estimates to avert risk. 
Retrospective analysis of portfolio performance 
with perfect earnings forecasts motivates this approach, 
demonstrating the superiority of LFM over standard factor approaches 
for both absolute returns and risk adjusted returns. 
In future work, we will examine how well the DNN forecasts 
compare to human analyst consensus forecasts 
and whether DNN forecasts can be improved 
by the consensus forecast via an ensemble approach. 
Finally, observing that there is a great amount 
of unstructured textual data about companies,
such as quarterly reports and earnings transcripts, 
we would like to explore whether such data 
can be used to improve our earnings forecasts.


\bibliographystyle{icml2020}
\bibliography{main}

\begin{thebibliography}{50}
\providecommand{\natexlab}[1]{#1}
\providecommand{\url}[1]{\texttt{#1}}
\expandafter\ifx\csname urlstyle\endcsname\relax
  \providecommand{\doi}[1]{doi: #1}\else
  \providecommand{\doi}{doi: \begingroup \urlstyle{rm}\Url}\fi

\bibitem[Bahdanau et~al.(2015)Bahdanau, Cho, and Bengio]{bahdanau2014neural}
Bahdanau, D., Cho, K., and Bengio, Y.
\newblock Neural machine translation by jointly learning to align and
  translate.
\newblock In \emph{International Conference on Learning Representations
  (ICLR)}, 2015.

\bibitem[Barclay \& Warner(1993)Barclay and Warner]{orderexec1993}
Barclay, M.~J. and Warner, J.~B.
\newblock {Stealth trading and volatility : Which trades move prices?}
\newblock \emph{Journal of Financial Economics}, 34\penalty0 (3):\penalty0
  281--305, December 1993.
\newblock URL
  \url{https://ideas.repec.org/a/eee/jfinec/v34y1993i3p281-305.html}.

\bibitem[Batres-Estrada(12015)]{batres2015deep}
Batres-Estrada, B.
\newblock Deep learning for multivariate financial time series.
\newblock Master's thesis, Royal Institute of Technology, Stockholm, Sweden,
  12015.

\bibitem[Bessembinder(2003)]{bessembinder_2003}
Bessembinder, H.
\newblock Trade execution costs and market quality after decimalization.
\newblock \emph{Journal of Financial and Quantitative Analysis}, 38\penalty0
  (4):\penalty0 747–777, 2003.
\newblock \doi{10.2307/4126742}.

\bibitem[Blundell(2017)]{Balaji2017}
Blundell, B. L. A. P.~C.
\newblock Simple and scalable predictive uncertainty estimation using deep
  ensembles.
\newblock \emph{Advances in Neural Information Processing Systems (NeurIPS)},
  2017.

\bibitem[Blundell et~al.(2015)Blundell, Cornebise, Kavukcuoglu, and
  Wierstra]{blundell2015weight}
Blundell, C., Cornebise, J., Kavukcuoglu, K., and Wierstra, D.
\newblock Weight uncertainty in neural networks.
\newblock \emph{International Conference on Machine Learning (ICML)}, 2015.

\bibitem[Boudoukh et~al.(2013)Boudoukh, Feldman, Kogan, and
  Richardson]{NBERw18725}
Boudoukh, J., Feldman, R., Kogan, S., and Richardson, M.
\newblock Which news moves stock prices? a textual analysis.
\newblock Working Paper 18725, National Bureau of Economic Research, January
  2013.
\newblock URL \url{http://www.nber.org/papers/w18725}.

\bibitem[Che et~al.(2018)Che, Purushotham, Cho, Sontag, and
  Liu]{che2016recurrent}
Che, Z., Purushotham, S., Cho, K., Sontag, D., and Liu, Y.
\newblock Recurrent neural networks for multivariate time series with missing
  values.
\newblock \emph{Scientific reports}, 8\penalty0 (1):\penalty0 1--12, 2018.

\bibitem[Chen et~al.(2015)Chen, Zhou, and Dai]{chen2015lstm}
Chen, K., Zhou, Y., and Dai, F.
\newblock A lstm-based method for stock returns prediction: A case study of
  china stock market.
\newblock In \emph{International Conference on Big Data (Big Data)}. IEEE,
  2015.

\bibitem[Cruz et~al.(2013)Cruz, Gomes, et~al.]{CRUZ2013}
Cruz, F. M.~d., Gomes, M. Y. F. S.~d., et~al.
\newblock The influence of rumors in the stock market: a case study with
  petrobras.
\newblock \emph{Transinforma{\c{c}}{\~a}o}, 25\penalty0 (3):\penalty0 187--193,
  2013.

\bibitem[D.~Zeiler(2012)]{adadelta}
D.~Zeiler, M.
\newblock Adadelta: An adaptive learning rate method.
\newblock \emph{arXiv:1212.570}, 1212, 2012.

\bibitem[Der~Kiureghian \& Ditlevsen(2009)Der~Kiureghian and
  Ditlevsen]{der2009aleatory}
Der~Kiureghian, A. and Ditlevsen, O.
\newblock Aleatory or epistemic? does it matter?
\newblock \emph{Structural safety}, 31\penalty0 (2):\penalty0 105--112, 2009.

\bibitem[Dimson et~al.(2017)Dimson, Marsh, and Staunton]{Dimson15}
Dimson, E., Marsh, P., and Staunton, M.
\newblock Factor-based investing: The long-term evidence.
\newblock \emph{The Journal of Portfolio Management}, 43\penalty0 (5):\penalty0
  15--37, 2017.
\newblock ISSN 0095-4918.
\newblock \doi{10.3905/jpm.2017.43.5.015}.
\newblock URL \url{https://jpm.iijournals.com/content/43/5/15}.

\bibitem[Ding et~al.(2015)Ding, Zhang, Liu, and Duan]{ding2015deep}
Ding, X., Zhang, Y., Liu, T., and Duan, J.
\newblock Deep learning for event-driven stock prediction.
\newblock In \emph{International Joint Conference on Artificial Intelligence
  (IJCAI)}, 2015.

\bibitem[Donahue et~al.(2015)Donahue, Anne~Hendricks, Guadarrama, Rohrbach,
  Venugopalan, Saenko, and Darrell]{donahue2015long}
Donahue, J., Anne~Hendricks, L., Guadarrama, S., Rohrbach, M., Venugopalan, S.,
  Saenko, K., and Darrell, T.
\newblock Long-term recurrent convolutional networks for visual recognition and
  description.
\newblock In \emph{Computer Vision and Pattern Recognition (CVPR)}, 2015.

\bibitem[Evgeniou \& Pontil(2004)Evgeniou and Pontil]{mtl1}
Evgeniou, T. and Pontil, M.
\newblock Regularized multi--task learning.
\newblock In \emph{Proceedings of the Tenth ACM SIGKDD International Conference
  on Knowledge Discovery and Data Mining}, KDD '04, pp.\  109--117, New York,
  NY, USA, 2004. ACM.
\newblock ISBN 1-58113-888-1.
\newblock \doi{10.1145/1014052.1014067}.
\newblock URL \url{http://doi.acm.org/10.1145/1014052.1014067}.

\bibitem[Fama \& French(1992)Fama and French]{fama1992returns}
Fama, E.~F. and French, K.~R.
\newblock The cross-section of expected stock returns.
\newblock \emph{Journal of Finance 47, 427-465.}, 1992.

\bibitem[Gal(2016)]{Gal2016Uncertainty}
Gal, Y.
\newblock \emph{Uncertainty in Deep Learning}.
\newblock PhD thesis, University of Cambridge, 2016.

\bibitem[Gal \& Ghahramani(2015{\natexlab{a}})Gal and
  Ghahramani]{Gal2015Dropout}
Gal, Y. and Ghahramani, Z.
\newblock Dropout as a {B}ayesian approximation: Insights and applications.
\newblock In \emph{International Conference on Machine Learning (ICML) Deep
  Learning Workshop}, 2015{\natexlab{a}}.

\bibitem[Gal \& Ghahramani(2015{\natexlab{b}})Gal and
  Ghahramani]{Gal2015DropoutC}
Gal, Y. and Ghahramani, Z.
\newblock Dropout as a bayesian approximation: Appendix.
\newblock \emph{arXiv:1506.02157}, 2015{\natexlab{b}}.

\bibitem[Gal et~al.(2017)Gal, Hron, and Kendall]{Gal2017Concrete}
Gal, Y., Hron, J., and Kendall, A.
\newblock {Concrete Dropout}.
\newblock \emph{Advances in Neural Information Processing Systems (NeurIPS)},
  2017.

\bibitem[Glorot \& Bengio(2010)Glorot and Bengio]{Glorot10}
Glorot, X. and Bengio, Y.
\newblock Understanding the difficulty of training deep feedforward neural
  networks.
\newblock In \emph{In Proceedings of the International Conference on Artificial
  Intelligence and Statistics (AISTATS’10). Society for Artificial
  Intelligence and Statistics}, 2010.

\bibitem[Goedhart et~al.(2005{\natexlab{a}})Goedhart, Koller, and
  Wessels]{mckinsey}
Goedhart, M., Koller, T., and Wessels, D.
\newblock Do fundamentals—or emotions—drive the stock market?
\newblock
  \url{https://www.mckinsey.com/business-functions/strategy-and-corporate-finance/our-insights/do-fundamentalsor-emotionsdrive-the-stock-market},
  2005{\natexlab{a}}.

\bibitem[Goedhart et~al.(2005{\natexlab{b}})Goedhart, Koller, and
  Wessels]{valuation}
Goedhart, M., Koller, T., and Wessels, D.
\newblock \emph{Valuation: Measuring and Managing the Value of Companies, 4th
  Edition}.
\newblock Wiley, 2005{\natexlab{b}}.

\bibitem[Goldberg(1989)]{Goldberg}
Goldberg, D.~E.
\newblock \emph{Genetic Algorithms in Search, Optimization and Machine
  Learning}.
\newblock Addison-Wesley Longman Publishing Co., Inc., Boston, MA, USA, 1st
  edition, 1989.
\newblock ISBN 0201157675.

\bibitem[Heskes(1997)]{Heskes97practicalconfidence}
Heskes, T.
\newblock Practical confidence and prediction intervals.
\newblock In \emph{Advances in Neural Information Processing Systems
  (NeurIPS)}. MIT press, 1997.

\bibitem[Hochreiter \& Schmidhuber(1997)Hochreiter and
  Schmidhuber]{hochreiter1997long}
Hochreiter, S. and Schmidhuber, J.
\newblock Long short-term memory.
\newblock \emph{Neural {C}omputation}, 9\penalty0 (8):\penalty0 1735--1780,
  1997.

\bibitem[Ioffe \& Szegedy(2015)Ioffe and Szegedy]{batchnorm}
Ioffe, S. and Szegedy, C.
\newblock Batch normalization: Accelerating deep network training by reducing
  internal covariate shift.
\newblock In \emph{International Conference on Machine Learning (ICML)}, 2015.

\bibitem[Jia(2016)]{jia2016investigation}
Jia, H.
\newblock Investigation into the effectiveness of long short term memory
  networks for stock price prediction.
\newblock \emph{arXiv:1603.07893}, 2016.

\bibitem[Kendall \& Gal(2017)Kendall and Gal]{kendall2017uncertainties}
Kendall, A. and Gal, Y.
\newblock What uncertainties do we need in bayesian deep learning for computer
  vision?
\newblock In \emph{Advances in neural information processing systems
  (NeurIPS)}, 2017.

\bibitem[{Khosravi} et~al.(2011){Khosravi}, {Nahavandi}, {Creighton}, and
  {Atiya}]{khosravi2011}
{Khosravi}, A., {Nahavandi}, S., {Creighton}, D., and {Atiya}, A.~F.
\newblock Comprehensive review of neural network-based prediction intervals and
  new advances.
\newblock \emph{IEEE Transactions on Neural Networks}, 22\penalty0
  (9):\penalty0 1341--1356, Sep. 2011.
\newblock ISSN 1045-9227.
\newblock \doi{10.1109/TNN.2011.2162110}.

\bibitem[Koku et~al.(1997)Koku, Jagpal, and Viswanath]{Koku1997}
Koku, P.~S., Jagpal, H.~S., and Viswanath, P.~V.
\newblock The effect of new product announcements and preannouncements on stock
  price.
\newblock \emph{Journal of Market-Focused Management}, 2\penalty0 (2):\penalty0
  183--199, Nov 1997.
\newblock ISSN 1572-8846.
\newblock \doi{10.1023/A:1009735620253}.
\newblock URL \url{https://doi.org/10.1023/A:1009735620253}.

\bibitem[Lipton et~al.(2016{\natexlab{a}})Lipton, Kale, Elkan, and
  Wetzell]{lipton2016learning}
Lipton, Z.~C., Kale, D.~C., Elkan, C., and Wetzell, R.
\newblock Learning to diagnose with lstm recurrent neural networks.
\newblock In \emph{International Conference on Learning Representations
  (ICLR)}, 2016{\natexlab{a}}.

\bibitem[Lipton et~al.(2016{\natexlab{b}})Lipton, Kale, and
  Wetzel]{lipton2016directly}
Lipton, Z.~C., Kale, D.~C., and Wetzel, R.
\newblock Directly modeling missing data in sequences with rnns: Improved
  classification of clinical time series.
\newblock \emph{Machine Learning for Healthcare (MLHC)}, 2016{\natexlab{b}}.

\bibitem[Mao et~al.(2015)Mao, Xu, Yang, Wang, Huang, and Yuille]{mao2014deep}
Mao, J., Xu, W., Yang, Y., Wang, J., Huang, Z., and Yuille, A.
\newblock Deep captioning with multimodal recurrent neural networks (m-rnn).
\newblock \emph{International Conference on Learning Representations (ICLR)},
  2015.

\bibitem[McGroarty et~al.(2018)McGroarty, Booth, Gerding, and
  Chinthalapati]{McGroarty2018}
McGroarty, F., Booth, A., Gerding, E., and Chinthalapati, V. L.~R.
\newblock High frequency trading strategies, market fragility and price spikes:
  an agent based model perspective.
\newblock \emph{Annals of Operations Research}, Aug 2018.
\newblock ISSN 1572-9338.
\newblock \doi{10.1007/s10479-018-3019-4}.
\newblock URL \url{https://doi.org/10.1007/s10479-018-3019-4}.

\bibitem[Menkveld(2016)]{hft2016}
Menkveld, A.~J.
\newblock The economics of high-frequency trading: Taking stock.
\newblock \emph{Annual Review of Financial Economics}, 8\penalty0 (1):\penalty0
  1--24, 2016.
\newblock \doi{10.1146/annurev-financial-121415-033010}.
\newblock URL \url{https://doi.org/10.1146/annurev-financial-121415-033010}.

\bibitem[Ng et~al.(2017)Ng, Gabriel, McAuley, Elkan, and
  Lipton]{ng2017predicting}
Ng, N., Gabriel, R.~A., McAuley, J., Elkan, C., and Lipton, Z.~C.
\newblock Predicting surgery duration with neural heteroscedastic regression.
\newblock In \emph{Machine Learning for Healthcare (MLHC)}, 2017.

\bibitem[{Nix} \& {Weigend}(1994){Nix} and {Weigend}]{NixWeig1994}
{Nix}, D.~A. and {Weigend}, A.~S.
\newblock Estimating the mean and variance of the target probability
  distribution.
\newblock In \emph{Proceedings of 1994 IEEE International Conference on Neural
  Networks (ICNN'94)}, volume~1, pp.\  55--60 vol.1, June 1994.
\newblock \doi{10.1109/ICNN.1994.374138}.

\bibitem[Pearce et~al.(2018)Pearce, Brintrup, Zaki, and
  Neely]{pmlr-v80-pearce18a}
Pearce, T., Brintrup, A., Zaki, M., and Neely, A.
\newblock High-quality prediction intervals for deep learning: A
  distribution-free, ensembled approach.
\newblock In Dy, J. and Krause, A. (eds.), \emph{Proceedings of the 35th
  International Conference on Machine Learning}, volume~80 of \emph{Proceedings
  of Machine Learning Research}, pp.\  4075--4084, Stockholmsmässan, Stockholm
  Sweden, 10--15 Jul 2018. PMLR.
\newblock URL \url{http://proceedings.mlr.press/v80/pearce18a.html}.

\bibitem[Pätäri \& Leivo(2017)Pätäri and Leivo]{Leivo2017value}
Pätäri, E. and Leivo, T.
\newblock A closer look at the value premium.
\newblock \emph{Journal of Economic Surveys, Vol. 31, Issue 1, pp. 79-168,
  2017}, 2017.

\bibitem[Qiu \& Song(2016)Qiu and Song]{qiu16}
Qiu, M. and Song, Y.
\newblock Predicting the direction of stock market index movement using an
  optimized artificial neural network model.
\newblock \emph{PLOS ONE}, 11:\penalty0 e0155133, 05 2016.
\newblock \doi{10.1371/journal.pone.0155133}.

\bibitem[Rothenstein et~al.(2011)Rothenstein, Tomlinson, Tannock, and
  Detsky]{oncology2011}
Rothenstein, J.~M., Tomlinson, G., Tannock, I.~F., and Detsky, A.~S.
\newblock {Company Stock Prices Before and After Public Announcements Related
  to Oncology Drugs}.
\newblock \emph{JNCI: Journal of the National Cancer Institute}, 103\penalty0
  (20):\penalty0 1507--1512, 09 2011.
\newblock ISSN 0027-8874.
\newblock \doi{10.1093/jnci/djr338}.
\newblock URL \url{https://doi.org/10.1093/jnci/djr338}.

\bibitem[Ruder(2017)]{mlt_seb}
Ruder, S.
\newblock An overview of multi-task learning in deep neural networks.
\newblock \emph{ArXiv}, abs/1706.05098, 2017.

\bibitem[Schumaker \& Maida(2018)Schumaker and Maida]{schumaker18}
Schumaker, R.~P. and Maida, N.
\newblock Analysis of stock price movement following financial news article
  release.
\newblock \emph{Communications of the IIMA}, 16, 2018.

\bibitem[Su et~al.(2018)Su, Ting, and Ansel]{EIM2018}
Su, D., Ting, Y.~Y., and Ansel, J.
\newblock Tight prediction intervals using expanded interval minimization.
\newblock \emph{CoRR}, abs/1806.11222, 2018.
\newblock URL \url{http://arxiv.org/abs/1806.11222}.

\bibitem[Sutskever et~al.(2014)Sutskever, Vinyals, and
  Le]{sutskever2014sequence}
Sutskever, I., Vinyals, O., and Le, Q.~V.
\newblock Sequence to sequence learning with neural networks.
\newblock In \emph{Advances in Neural Information Processing (NeurIPS)}, 2014.

\bibitem[Tripathi et~al.(2016)Tripathi, Lipton, Belongie, and
  Nguyen]{tripathi2016context}
Tripathi, S., Lipton, Z.~C., Belongie, S., and Nguyen, T.
\newblock Context matters: Refining object detection in video with recurrent
  neural networks.
\newblock In \emph{British Machine Vision Conference (BMVC)}, 2016.

\bibitem[Vinyals et~al.(2015)Vinyals, Toshev, Bengio, and
  Erhan]{vinyals2015show}
Vinyals, O., Toshev, A., Bengio, S., and Erhan, D.
\newblock Show and tell: A neural image caption generator.
\newblock In \emph{Computer Vision and Pattern Recognition (CVPR)}, 2015.

\bibitem[Wanjawa \& Muchemi(2014)Wanjawa and Muchemi]{wanjawa2014ann}
Wanjawa, B.~W. and Muchemi, L.
\newblock Ann model to predict stock prices at stock exchange markets.
\newblock \emph{arXiv:1502.06434}, 2014.

\end{thebibliography}

\end{document}